\documentclass[11pt,twoside,letterpaper]{article}
\usepackage{pslatex}
\usepackage{fancyhdr}
\usepackage{amsthm}
\usepackage[dvips]{graphicx}
\usepackage{geometry}
\usepackage{times}
\usepackage{amssymb}
\usepackage{amsmath}
\usepackage{makeidx}

\makeatletter
\def\cleardoublepage{\clearpage\if@twoside \ifodd\c@page\else    \hbox{}    \thispagestyle{empty}    \newpage    \if@twocolumn\hbox{}\newpage\fi\fi\fi}
\makeatother

\def\figurename{Figure}
\makeatletter
\renewcommand{\fnum@figure}[1]{\figurename~\thefigure.}
\makeatother
\def\tablename{Table}
\makeatletter
\renewcommand{\fnum@table}[1]{\tablename~\thetable.}
\makeatother

\input{tcilatex}

\begin{document}

\title{ {The Principle of Solidarity: Geometrical Description of Interactions}}
\author{\bfseries\itshape Fabio Cardone$^{1,2}$, Roberto Mignani$^{2-4}$ and
Andrea Petrucci$^{3}$ \\
%EndAName
$^{1}$\,Istituto per lo Studio dei Materiali Nanostrutturati (ISMN -- CNR)\\
Via dei Taurini - 00185 Roma, Italy\\
$^{2}$\,GNFM, Istituto Nazionale di Alta Matematica ``F.Severi''\\
\ Citt\`{a} Universitaria, P.le A.Moro 2 - 00185 Roma, Italy\\
$^{3}$\,Dipartimento di Fisica ``E. Amaldi'', \\
Universit\`{a} degli Studi ``Roma Tre'' \\
\ Via della Vasca Navale, 84 - 00146 Roma, Italy\\
$^{4}$\,I.N.F.N. - Sezione di Roma III }
\date{}
\maketitle

\begin{abstract}
We discuss the possibility of geometrizing interactions by exploiting the
``principle of solidarity'' between space-time and the \ physical phenomena
occurring in it (formulated by the Italian matematician B. Finzi in 1955).
This is accomplished \ by means of a deformation of the Minkowski metric,
implemented by assuming that the metric coefficients \ depend on \ the
energy of the process considered. Such a formalism (``Deformed Special \
Relativity'') allows one, among the others, to deal with the \ breakdown of
Lorentz \ invariance and to recover it in a generalized sense.
\end{abstract}

\thispagestyle{empty} \setcounter{page}{19}
% ------- [First Page Running Head] - place it immediately after title! ------
\thispagestyle{fancy} \fancyhead{}
\fancyhead[L]{In: Einstein and Hilbert: Dark Matter \\
Editor: V. V. Dvoeglazov, pp. {\thepage-\pageref{lastpage-01}}}
% needs \label{lastpage-01} on the last page.
\fancyhead[R]{ISBN 978-1-61324-840-9  \\
\copyright~2011 Nova Science Publishers, Inc.} \fancyfoot{} %
\renewcommand{\headrulewidth}{0pt}
%------------------------------------------------------------------------------

\section{Introduction: The Finzi Principle of Solidarity}

In 1955 the Italian mathematician Bruno Finzi, in his contribution to the
book \textit{``Fifty Years of Relativity''}[1]\textit{, }stated his \emph{%
``Principle of Solidarity''}(PS)\footnote{{It's quite difficult to express
in English in a simple way the Italian words ``\textit{solidariet\`{a}}''
and ``\textit{solidale}'', used by Finzi to mean the feedback between
space-time and interactions. A possible way to render them is to use ``%
\textit{solidarity}'' and ``\textit{solidly connected}'', respectively - at
the price of partially loosing the common root of the Italian words - , with
the warning that what Finzi really means is that the very structure of
space-time is determined by the physical phenomena which do take place in it.%
}}, that sounds ``\textit{It's (indeed) necessary to consider space-time TO\
BE SOLIDLY CONNECTED with the physical phenomena occurring in it, so that
its features and its very nature do change with the features and the nature
of those. In this way not only (as in classical and special-relativistic
physics)\ space-time properties affect phenomena, but reciprocally phenomena
do affect space-time properties. One thus recognizes in such an appealing
``Principle of Solidarity'' between phenomena and space-time that
characteristic of mutual dependence between entities, which is peculiar to
modern science.}'' Moreover, referring to a generic N-dimensional space:
\textit{``It can, }a priori\textit{, be pseudoeuclidean, Riemannian,
non-Riemannian. But} --- he wonders --- \textit{how is indeed the space-time
where physical phenomena take place? Pseudoeuclidean, Riemannian,
non-Riemannian, according to their nature, as requested by the principle of
solidarity between space-time and phenomena occurring in it.''}

% ------------ [Running Heads - for odd and even pages] - please insert it only on page 2!
\pagestyle{fancy} \fancyhead{} \fancyhead[EC]{Fabio Cardone, Roberto Mignani
and Andrea Petrucci} \fancyhead[EL,OR]{\thepage} \fancyhead[OC]{The
Principle of Solidarity: Geometrizing Interactions} \fancyfoot{} %
\renewcommand\headrulewidth{0.5pt}
%------------------------------------------------------------------------------

Of course, Finzi's main purpose was to apply such a principle to Einstein's
Theory of General Relativity, namely to the class of gravitational
phenomena. However, its formulation is as general as possible, so to apply
in principle to all the known physical interactions. Therefore, Finzi's PS
is at the very ground of any attempt at geometrizing physics, i.e.
describing physical forces in terms of the geometrical structure of
space-time.

Such a project (pioneered by Einstein himself) revealed itself unsuccessful
even when only two interactions were known, the electromagnetic and the
gravitational one. It was fully abandoned starting from the middle of the
XXth century, due to the discovery of the two nuclear interactions, the weak
and the strong one (apart from recent attempts based on string theory).

The basic problem is how to implement Finzi's Principle of Solidarity for
all interactions on a mere geometrical basis. Since, from an historical
point of view, General Relativity (GR) is the only successful theoretical
realization of geometrizing an interaction (the gravitational one), it is
usually believed that the goal of geometrization of interactions can only be
achieved by the tools of Riemannian spaces or of their suitable
generalizations.

We want instead to show that implementing the Finzi principle can be
obtained in the mere framework of Special Relativity, provided its very
foundations are taken into proper account and suitably exploited. To this
aim, let us analyze Special Relativity from an axiomatic standpoint.

\section{An Axiomatic View to Special Relativity}

Special Relativity (SR) is essentially grounded on the properties of
space-time, \textit{i.e.} isotropy of space and homogeneity of space and
time \ (as a consequence of the equivalence of inertial frames) and on the
principle of relativity.

The two basic postulates of SR in its axiomatic formulation are [2]:

\noindent 1 -\emph{\ Space-time properties}: Space and time are homogeneous
and space is isotropic.

\noindent 2 - \emph{Principle of Relativity (PR)}: All physical laws must be
covariant when passing from an inertial reference frame $K$ to another frame
$K^{\prime }$, moving with constant velocity relative to $K$.

The second postulate can be traced back to Galilei himself, who of course
enunciated and applied it with reference to the laws of classical mechanics
(the only ones known at his times). In fact, the Relativity Principle
contains implicitly (somewhat hidden, but actually easily understood after a
moment's thought) the basic point that, for a correct formulation of SR,
\emph{it is\ necessary to specify the total class, }$C_{T}$\emph{, of the
physical phenomena to which the PR applies. }The importance of such a
specification is easily seen if one thinks that, from an axiomatic
viewpoint, the only difference between Galilean and Einsteinian relativities
just consists in the choice of $C_{T}$ (\textit{i.e.} the class of
mechanical phenomena in the former case, and of mechanical and
electromagnetic phenomena in the latter).

It is possible to show that, from the above two postulates, there follow ---
without any additional hypothesis --- all the usual ``principles'' of SR,
\textit{i.e.} the ``principle of reciprocity'', the linearity of
transformations between inertial frames, and the invariance of light speed
in vacuum.

Concerning this last point, it can be shown in general that postulates 1 and
2 above imply the existence of an invariant, real quantity, having the
dimensions of the square of a speed, whose value must be experimentally
determined in the framework of the total class $C_{T}$ of the physical
phenomena\footnote{{The invariant speed is obviously $\infty $ for Galilei's
relativity, and $c$ (light speed in vacuum) for Einstein's relativity.}}.
Such an invariant speed depends on the interaction (fundamental, or at least
phenomenological) ruling the physical phenomenon considered. Therefore \emph{%
there is, a priori, an invariant speed for every interaction}, namely, a
maximal causal speed for every interaction.

All the formal machinery of SR in the Einsteinian sense (including Lorentz
transformations and their implications, and the metric structure of
space-time) is simply a consequence of the above two postulates and of the
choice, for the total class of physical phenomena $C_{T}$, of the class of
mechanical and electromagnetic phenomena.

If different explicit choices of $C_{T}$ are made, one gets a priori
different realizations of the theory of relativity (in its abstract sense),
each one embedded in the previous. Of course, the principle of relativity,
together with the specification of the total class of phenomena considered,
necessarily entails in all cases, for consistency, the uniqueness of the
transformation equations connecting inertial reference frames\footnote{%
The hypothesis of the existence \textit{a priori }of different relativities
for different interactions --- formulated by Recami and one of the present
authors (R.M.) on the basis of the above critical analysis of the
foundations of Special Relativity --- can be considered a generalization of
the point of view advocated by Lorentz, according to which different
interactions require different coordinate transformations between inertial
reference frames.}.

The attempt at including the class of nuclear and subnuclear phenomena in
the total class of phenomena for which Special Relativity holds true is
therefore expected to imply a generalization of the Minkowski metric,
analogously to the generalization from the Euclidean to the Minkowski metric
in going from mechanics to electrodynamics.

However, in order to avoid misunderstandings, it must be stressed that such
an analogy with the extension of the Euclidean metric has to be understood
not in the purely geometric meaning, but rather in the sense (as already
stressed by Penrose [3]) of Euclidean geometry as a physical theory\textit{.
}

Indeed, the generalized metric must be equipped with a dynamic character and
be not only a consequence, but also an effective description of (the
interaction involved in) the class of phenomena considered. This allows one
in this way to get a feedback between interactions and space-time structure,
already accomplished for gravitation in General Relativity.

This complies with the \emph{''Principle of Solidarity''} stated by Finzi in
the form already quoted above, which can be embodied in the following third
principle of Relativity: \newline
3 - \emph{Principle of Solidarity (PS)}: Each class of phenomena (namely,
each interaction) determines its own space-time.

The fundamental problem is now: \emph{How to endow the metric of the
Min\-kowski space-time with a geometrical structure able to describe the
interaction involved in a given process?} We will answer this question in
the following.

\section{Energy and the Finzi Principle}

At present, General Relativity (GR) is the only successful theoretical
realization of geometrizing an interaction (the gravitational one). As is
well known, energy plays a fundamental role in GR, since the energy-momentum
tensor of a given system is the very source of the gravitational field.

A moment's thought shows that this occurs actually also for other
interactions. Let us remind, for instance, the case of Euclidean geometry in
its intrinsic meaning of a theory of physical reality at its basic classical
(macroscopic) level. In fact, it describes in a quantitative way, in
mathematical language, the relations among measured physical entities ---
distances, in this case ---, and therefore the physical space in which
phenomena occur.

However, the measurement of distances depends on the motion of the body
which actually performs the measurement. Such a dependence is indeed not on
the \emph{kind }of motion, but rather on the \emph{energy }needed to let the
body move, and on the \emph{interaction} providing such energy. The
measurement of time needs as well a periodic motion with constant frequency,
and therefore it too depends on the energy and on the interaction.

This simple example shows how \emph{energy does play a fundamental role in
determining the very geometrical structure of space-time }(in analogy\emph{\
}with the General-Relativistic case, where --- as already noted --- the
energy-momentum tensor is the source of the gravitational field). Let us
stress that such a viewpoint is very similar, on many respects, to the
Ehlers-Pirani-Schild scheme [4] (based on the earlier work of Weyl), in
which the geometry of space-time is operationally determined by using the
trajectories of free-falling objects (geodesics). In this framework, the
points of space-time become physically real in virtue of the geometrical
relations between them, and the classical particle motion is exploited to
obtain the geometry of space-time (the argument can be extended to quantum
motion as well [5]).

Generalizing such an argument, we can state that exchanging energy between
particles amounts to measure operationally their space-time separation. Of
course such a process depends on the interaction involved in the energy
exchange; moreover, each exchange occurs at the maximal causal speed
characteristic of the given interaction. It is therefore natural to assume
that the measurement of distances, performed by the energy exchange
according to a given interaction, realizes the ``solidarity principle''
between space-time and interactions at the microscopic scale\textbf{.}

By starting from such considerations, a possible way to implement Finzi's
principle for \emph{all} fundamental interactions is provided by the
formalism of \emph{Deformed Special Relativity }(\emph{DSR}) developed in
the last decade of the XX century. It is based on a \emph{deformation} of
the Minkowski space, namely a space-time endowed with a metric whose
coefficients just depend on energy (in the sense specified later on). Such
an energy-dependent metric does assume a \emph{dynamic role}, thus providing
a geometrical description of the fundamental interaction considered and
implementing the feedback between space-time structure and physical
interactions which is just the content and the heritage of Finzi's principle.

The generalization of the Minkowski space implies, among the others, new,
generalized transformation laws, which admit, as a suitable limit, the
Lorentz transformations (just like Lorentz transformations represent a
covering of the Galilei-Newton transformations) [6].

Then, the solidarity principle allows one to recover the basic features of
the relativity theory in the Lorentz (not Einstein) view (Lorentzian
relativity), namely different interactions entail different coordinate
transformations and different invariant speeds.

\section{\textbf{Description of Interactions by Energy-Dependent \newline
Metrics}}

We will now show how the dynamic role of the energy, in describing the
structure of space-time, can be exploited in order to geometrize all four
fundamental interactions, so to comply with the Finzi principle. As already
stressed above, this can be achieved by suitably \emph{deforming}
space-time, according to what dictated by the energy involved in the
process, ruled by the interaction considered. Speaking in a figurative
language, we can say that in such a view spacetime is not a rigid (and
passive) background, but a sort of elastic carpet, able to change its shape
according to the (energy of) the interaction involved, and to react in turn
on the process, thus affecting its dynamics in a active way.

\subsection{Deformed Minkowski Space-Time}

In the attempt at a geometrical implementation of the Finzi principle, we
have therefore to take into account the role of energy in determining an
interaction, and the different ``relativities'' obtained in correspondence
to different classes of physical phenomena.

As is well known, the Minkowski metric \footnote{%
In the following, lower Latin indices take the values $\left\{ 1,2,3\right\}
$\ and label spatial dimensions, whereas lower Greek indices vary in the
range $\left\{ 0,1,2,3\right\} $, with 0 referring to the time dimension.
For brevity's sake, we shall denote simply by $x$ the (contravariant)
four-vector ($x^{0},x^{1},x^{2},x^{3}$). Moreover, we adopt the signature $%
\left( +,-,-,-\right) $ for the four-dimensional spacetime, and employ the
notation ``$ESC$ $on$''\ (``$ESC$ $off$'') to mean that the Einstein sum
convention on repeated indices is (is not) used.}
\begin{equation}
g=diag(1,-1,-1,-1)
\end{equation}%
is a generalization of the Euclidean metric $\epsilon =diag(1,1,1)$. By the
considerations of the previous sections, we can assume that the metric
describes, in an effective way, the interaction, and that there exist
interactions more general than the electromagnetic ones (which, as well
known, are long-range and derivable from a potential).

The simplest generalization of the space-time metric which accounts for such
more general properties of interactions is provided by a \emph{deformation},
$\eta $, of the Minkowski metric (1), defined as [6]
\begin{equation}
\eta =diag(b_{0}^{2},-b_{1}^{2},-b_{2}^{2},-b_{3}^{2}).
\end{equation}

Of course, from a formal point of view metric (2) is not new at all.
Deformed Minkowski metrics of the same type have already been proposed in
the past in various physical frameworks, starting from Finsler's
generalization of Riemannian geometry [7] to Bogoslowski's anisotropic
space-time [8] (just based on a Finslerian metric) to the isotopic Minkowski
space [9]. A phenomenological deformation of the type (2) was also obtained
by Nielsen and Picek [10] in the context of the electroweak theory.
Moreover, although for quite different purposes, \textquotedblleft
quantum\textquotedblright\ deformed Minkowski spaces have been also
considered in the context of quantum groups [11]. Leaving to later
considerations the true specification of the exact meaning of the deformed
metric (2) in our framework, let us right now stress two basic
points.\medskip

1 - Firstly, metric (2) is supposed to hold at a\emph{\ local} (and not
global) scale, \textit{i.e.} to be valid not everywhere, but only in a
suitable (local) space-time region (characteristic of both the system and
the interaction considered). We shall therefore refer often to it as a ``%
\emph{topical}'' deformed metric\footnote{%
Notice that the assumed local validity of (2) differentiates this approach
from those based on Finsler's geometry or from the Bogoslowski's one (which,
at least in their standard meaning, do consider deformed metrics at a\emph{\
global} scale), and makes it similar, on some aspects, to the philosophy and
methods of the isotopic generalizations of Minkowski spaces [9]. However, it
is well known that Lie-isotopic theories rely in an essential way, from the
mathematical standpoint, on (and are strictly characterized by) the very
existence of the so-called isotopic unit. In the following, such a formal
device will not be exploited (because unessential on all respects), so that
the present formalism is not an isotopic one. Moreover, from a physical
point of view, the isotopic formalism is expected to apply only to strong
interactions. On the contrary, it will be assumed that the (effective)
representation of interactions through the deformed metric (2) does hold for
\emph{all} kinds of interactions (at least for their nonlocal component). In
spite of such basic differences this formalism shares some common formal
results --- as we shall see in the following --- with isotopic relativity
(like the mathematical expression of the generalized Lorentz
transformations: see [6).]}.

In the present case, the term `local'' must be understood in the sense that
a deformed metric of the kind (2) describes the geometry of a 4-dimensional
variety attached at a point $x$ of the standard Minkowski space-time, in the
same way as a local Lorentz frame is associated (as a tangent space) to each
point of the (globally Riemannian) space of Einstein's GR. Another example,
on some respects more similar to the present formalism, is provided by a
space-time endowed with a vector fibre-bundle structure, where a Riemann
space with constant curvature is attached at each point $x$.\smallskip

2 - Secondly, metric (2) is regarded to play a \emph{dynamic role}. So, in
order to comply with the solidarity principle, we assume that the parameters
$b_{\mu }(\mu =0,1,2,3)$ are, in general, real and positive functions of a
given set of observables $\left\{ \mathcal{O}\right\} $ characterizing the
system (in particular, of its total energy exchange, as specified later):
\begin{equation}
\left\{ b_{\mu }\right\} =\left\{ b_{\mu }(\left\{ \mathcal{O}\right\}
)\right\} \in R_{0}^{+},\text{ }\forall \text{{\footnotesize \ }}\left\{
\mathcal{O}\right\} \text{{\footnotesize \ }}
\end{equation}%
The set $\left\{ \mathcal{O}\right\} $ represents therefore, in general, a
set of non-metric variables ($\left\{ x_{n.m.}\right\} $).

Eq.~(2) therefore becomes:
\begin{gather}
\eta _{\mu \nu }=\eta _{\mu \nu }(\left\{ \mathcal{O}\right\} )\smallskip
\notag \\
=diag(b_{0}^{2}(\left\{ \mathcal{O}\right\} ),-b_{1}^{2}(\left\{ \mathcal{O}%
\right\} ),-b_{2}^{2}(\left\{ \mathcal{O}\right\} ),-b_{3}^{2}(\left\{
\mathcal{O}\right\} )).\smallskip
\end{gather}

However, for the moment the deformation of the Minkowski space will be
discussed only from a formal point of view, by disregarding the problem of
the observables on which the coefficients $b_{\mu }$ actually depend (it
will be faced later on).

It is now possible to define a generalized (\emph{``deformed''}) Minkowski
space $\widetilde{M}(x,\eta (\left\{ \mathcal{O}\right\} ))$ with the same
local coordinates $x$ of $M$ (the four-vectors of the usual Minkowski
space), but with metric given by the metric tensor $\eta $ (4). The
generalized interval in $\widetilde{M}$ is therefore given by $(x^{\mu
}=(x^{0},x^{1},x^{2},x^{3})=(ct,x,y,z)$, with $c$ being the usual light
speed in vacuum) (ESC on) [6]:
\begin{gather}
ds^{\tilde{2}}(\left\{ \mathcal{O}\right\} )  \notag \\
\equiv b_{0}^{2}(\left\{ \mathcal{O}\right\} )c^{2}dt^{2}-b_{1}^{2}(\left\{
\mathcal{O}\right\} )\text{(}dx^{1}\text{)}^{2}-b_{2}^{2}(\left\{ \mathcal{O}%
\right\} )\text{(}dx^{2}\text{)}^{2}-b_{3}^{2}(\left\{ \mathcal{O}\right\} )%
\text{(}dx^{3}\text{)}^{2}  \notag \\
=\eta _{\mu \nu }(\left\{ \mathcal{O}\right\} )dx^{\mu }dx^{\nu }=dx\ast dx.
\end{gather}

The last step in (5) defines the scalar product $\ast $ in the deformed
Minkowski space $\widetilde{M}$.

It is worth to recall that the deformation of the metric, resulting in the
interval (5), represents a geometrization of a suitable space-time region
(corresponding to the physical system considered) that describes, in the
average, the effect of nonlocal interactions on a test particle. It is clear
that there exist infinitely many deformations of the Minkowski space
(precisely, $\infty ^{4}$), corresponding to the different possible choices
of the parameters $b_{\mu }$, a priori different for each physical system.

Moreover, since the usual, ``flat'' Minkowski metric $g$ (1) is related in
an essential way to the electromagnetic interaction, it must be understood
that electromagnetic interactions imply the presence of a fully Minkowskian
metric\footnote{{Actually, a deformed metric of the type (4) is required if
one wants to account for possible nonlocal electromagnetic effects (see [6]).%
}}.

Once the mathematical body of our formalism is specified, one has now to
give a physical soul to it, in order to comply with the Finzi principle. On
the basis of the discussion of Sect.3, \emph{we have to take, as observable }%
$\mathcal{O}$ \emph{on which the metric coefficients }$b_{\mu }(\left\{
\mathcal{O}\right\} )$\ \emph{depend, the total energy }$E$\emph{\ exchanged
by the physical system considered during the interaction process:}

\begin{equation}
\left\{ \mathcal{O}\right\} \equiv E\overset{}{\Leftrightarrow }\left\{
b_{\mu }(\left\{ \mathcal{O}\right\} )\right\} \equiv \left\{ b_{\mu
}(E)\right\} ,\quad \forall \mu =0,1,2,3.
\end{equation}%
Actually, since all the functions $\left\{ b_{\mu }\right\} $ are
dimensionless, they must depend on a dimensionless variable. Then, one has
to divide the energy $E$ by a constant $E_{0}$ (in general characteristic of
each fundamental interaction), with dimensions of energy, so that:
\begin{equation}
\left\{ b_{\mu }(\left\{ \mathcal{O}\right\} )\right\} \equiv \left\{ b_{\mu
}(E/E_{0})\right\} ,\quad \forall \mu =0,1,2,3.
\end{equation}

Thus, the distance measurement is accomplished by means of the deformed
metric tensor function of the energy, given explicitly by
\begin{equation}
\eta _{\mu \nu }(E)=diag(b_{0}^{2}\left( E\right)
,-b_{1}^{2}(E),-b_{2}^{2}(E),-b_{3}^{2}(E)).  \notag
\end{equation}

Any interaction can be therefore phenomenologically described by metric (8)
in an \emph{effective} way. This is true in general, but necessary in the
case of nonlocal and nonpotential interactions. For force fields which admit
a potential, such a description is complementary to the actual one.

One is therefore led to put forward a revision of the concept of
``geometri\-zation of an interaction'': each interaction produces its own
metric, formally expressed by the metric tensor (8), but realized via
different choices of the set of parameters $b_{\mu }(E) $. Otherwise said,
the $b_{\mu }(E)$'s are peculiar to every given interaction. The statement
that (8) provides us with a metric description of an interaction must be
just understood in such a sense.

Therefore, the energy-dependent deformation of the Minkowski metric
implements a generalization of the concept of geometrization of an
interaction (in accordance with Finzi's principle). The GR theory implements
a geometrization (at a \emph{global } scale) of the gravitational
interaction, based on its derivability from a potential and on the
equivalence between the inertial mass of a body and its ``gravitational
charge''. The formalism of energy-dependent metrics allows one instead to
implement a geometrization (at a \emph{local } scale) of any kind of
interaction, at least on a phenomenological basis. As already stressed
before, such a formalism applies, in principle, to both fundamental and
phenomenological interactions, either potential (gravitational,
electromagnetic) or nonpotential (strong, weak), \emph{local} and \emph{%
nonlocal}, for which either an Equivalence Principle holds (as it is the
case of gravitation) or (in the more general case) the inertial mass of the
body \emph{is not} in general proportional to its charge in the force field
considered (e.m., strong, and weak interaction).

Let us explicitly stress that the theory of SR based on metric (4) has
nothing to do with General Relativity. Indeed, in spite of the formal
similarity between the interval (5), with the $b_{\mu }$ functions of the
coordinates, and the metric structure of a Riemann space, in this framework
no mention at all is made of the equivalence principle between mass and
inertia, and among non-inertial, accelerated frames. Moreover, General
Relativity describes geometrization on a large-scale basis, whereas the
special relativity with topical deformed metric describes local
(small-scale) deformations of the metric structure (although the term
``small scale'' must be referred to the real dimensions of the physical
system considered). But the basic difference is provided by the fact that
actually the deformed Minkowski space $\widetilde{M}$\ has zero curvature,
as it is easily seen by remembering that, in a Riemann space, the scalar
curvature is constructed from the derivatives, with respect to space-time
coordinates, of the metric tensor. In others words, the space $\tilde{M}$ is
\emph{intrinsically flat }--- at least in a mathematical sense.

Namely, it would be possible, in principle, to find a change of coordinates,
or a rescaling of the lengths, so as to recover the usual Minkowski space.
However, such a possibility is only a mathematical, and not a physical one.
This is related to the fact that the energy of the process is fixed, and
cannot be changed at will. For that value of the energy, the metric
coefficients do possess values different from unity, so that the
corresponding space $\tilde{M}$, for the given energy value, is actually
different from the Minkowski one. The usual space-time $M$\ is recovered for
a special value $E_{0}$\ of the energy (characteristic of any interaction),
such that indeed
\begin{equation}
\eta (E_{0})=g=diag(1,-1,-1,-1).
\end{equation}%
Such a value $E_{0}$ (which must be derived from the phenomenology) will be
referred to as \emph{the threshold energy of the interaction considered}. It
can be seen that it is just the constant appearing in Eq.~(7).

\subsection{Energy as Dynamic Variable}

The basic point of the present way of geometrizing an interaction (thus
implementing the Finzi legacy) consists in an \textbf{``}upsetting\textbf{''}
of the space-time-energy parametrization. Whereas for potential interactions
there exists a potential energy depending on the space-time metric
coordinates, one has here to deal with a deformed metric tensor $\eta $,
whose coefficients depend on the energy, that thus assumes a \emph{dynamic }%
role. However, the identification of energy as the physical observable on
which the metric must depend\ leaves open the question, what energy? Let us
answer this question.

From the physical point of view, $E$ is the measured energy of the system,
and thus a merely phenomenological variable. As is well known, all the
present physically realizable detectors work \textit{via} their
electromagnetic interaction in the usual space-time $M$. This is why, in
this formalism, the Minkowski space and the e.m. interaction do play a
fundamental role. The former is --- as already stressed --- the cornerstone
on which to build up the generalization of Special Relativity based on the
deformed metric (8). The latter is the comparison term for all fundamental
interactions. Let us recall that they are strictly interrelated, since it is
just electromagnetism which determines the Minkowski geometry. Then, stating
that the measurement of $E$ occurs \textit{via} the e.m. interaction amounts
to say that it is measured in $M$. This ensures that the total energy is
conserved, due the validity of the Hamilton theorem in Minkowski space. In
summary,\emph{\ }$E$\emph{\ has to be understood as the energy measured by
the detectors through the e.m. interaction in Minkowskian conditions and
under validity of total energy conservation.}

From the mathematical standpoint, $E$ has to be considered as a dynamic
variable, because it specifies the dynamic behavior of the process under
consideration, and, through the metric coefficients, provides us with a
dynamic map --- in the energy range of interest --- of the interaction
ruling the given process.

Let us notice that metric (8) plays, for nonpotential interactions, a role
analogous to that of the Hamiltonian $H$ for a potential interaction. In
particular, the metric tensor $\eta $ as well is not an input of the theory,
but must be built up from the experimental knowledge of the physical data of
the system concerned (in analogy with the specification of the Hamiltonian
of a potential system). However, there are some differences between $\eta $
and $H$ worth to be stressed. Indeed, as is well known, $H$ represents the
total energy $E_{tot}$ of the system irrespective of the value of $E_{tot}$
and the choice of the variables. On the contrary, $\eta (E)$ describes the
variation in the measurements of space and time, in the physical system
considered, as $E_{tot}$ changes; therefore, $\eta $ does depend on the
numerical value of $H$, but not on its functional form. The explicit
expression of $\eta $ depends only on the interaction involved\footnote{%
It is worth recalling that the use of an energy-dependent space-time metric
can be traced back to Einstein himself, who generalized the Minkowski
interval as follows
\par
\begin{equation*}
ds^{2}=\left( 1+\frac{2\phi }{c^{2}}\right)
c^{2}dt^{2}-(dx^{2}+dy^{2}+dz^{2})
\end{equation*}%
(where $\phi $ is the Newtonian gravitational potential), in order to
account for the modified rate of a clock in presence of a (weak)
gravitational field.}.

One may be puzzled about the dependence of the metric on the energy, which
is not an invariant under usual Lorentz transformations, but transforms like
the time-component of a four vector.

Actually, energy has to be regarded, in this formalism, from two different
points of view. One has, on one side, the energy as measured in full
Minkowskian conditions, which, as such, behaves as a genuine four-vector%
\emph{\ }under usual Lorentz transformations\textit{\ }(in the sense that it
changes in the usual way if we go, say, from the laboratory frame to another
frame in uniform motion with respect to it). Once fixed the frame, one gets\
a measured value of the energy for a given process. This is the value which
enters,\emph{\ }as a parameter\textit{,} in the expression (8) of the
deformed metric. Such an energy, therefore, is no longer to be considered as
a four vector in the deformed Minkowski space\textit{, }but it is just a
quantity whose value determines the deformed geometry of the process
considered (or, otherwise speaking, which selects the deformed space-time we
have to use to describe the phenomenon)\footnote{%
This different view to energy constitutes the basic point to building up a
five-dimensional space-time, in which $E$ does just represent the extra
dimension (see [6]).}.

The problem of a metric description of a given interaction is thus formally
reduced to the determination of the coefficients $b_{\mu }(E)$ from the data
on some physical system, whose dynamic behavior is ruled by the interaction
considered.

\subsection{Deformed Special Relativity}

In order to develop the relativity theory in a deformed Minkowski
space-time, one has to suitably generalize and clarify the basic concepts
which are at the very foundation of SR.

Let us first of all define a ``topical inertial frame'':\smallskip

\textbf{i)} A \emph{topical ''inertial'' frame} (TIF) is a reference frame
in which space-time is homogeneous, but space is not necessarily
isotropic.\smallskip

Then, a \emph{``generalized principle of relativity''}, or \emph{``principle
of metric invariance''}, can be stated as follows:\smallskip

\textbf{ii)} all physical measurements within every topical ''inertial''
frame must be carried out via the \emph{same} metric.\smallskip

We named \emph{``Deformed Special Relativity''} (DSR) [6] the generalization
of SR based on the above two postulates, and whose space-time structure is
given by the deformed Minkowski space $\widetilde{M}$ introduced in Sect. 2.
Let us also warn the reader against confusing this formalism with a
different generalization of SR, i.e. Doubly Special Relativity [12], that
uses the same acronym. This latter theory is essentially based on the
quantum deformation of the Poincar\'{e} algebra, precisely, its $\varkappa $%
-deformation. In such a kind of deformation, one essentially modifies the
commutation relations of the Poincar\'{e} generators, whereas in the DSR
framework the deformation concerns primarily the metrical structure of the
space-time (although the Poincar\'{e} algebra is affected, too: see [6]).
However, it is not clear at present if the two theories may have some points
in common (for instance, the energy dependence of the metric in position
space).

Moreover, henceforth we shall use the notation $g_{DSR}$ for the metric
tensor of DSR (in order to distinguish it from --- but also to stress its
affinities with --- the standard Minkowskian metric tensor $g\equiv g_{SR}$%
), so that (with reference to Eq.~(8))
\begin{equation}
g_{\mu \nu ,DSR}(E)=diag(b_{0}^{2}\left( E\right)
,-b_{1}^{2}(E),-b_{2}^{2}(E),-b_{3}^{2}(E))
\end{equation}%
is the covariant\emph{\ }deformed metric tensor of $\widetilde{M}$.

The corresponding deformed interval is of course
\begin{gather}
ds^{\tilde{2}}(E)=g_{\mu \nu ,DSR}(E)dx^{\mu }dx^{\nu }  \notag \\
=b_{0}^{2}(E)c^{2}dt^{2}-b_{1}^{2}(E)dx^{2}-b_{2}^{2}(E)dy^{2}-b_{3}^{2}(E)dz^{2}.
\end{gather}

In matrix notation, the deformed interval (11) reads
\begin{equation}
ds^{\tilde{2}}(E)=\left( dX\right) ^{T}g_{DSR}(E)dX,
\end{equation}%
where $dX$ is the $4\times 1$ column vector with elements $dx^{\mu }$\ (so
that $\left( dX\right) ^{T}=(dx^{0}$ $dx^{1}$ $dx^{2}$ $dx^{3})$, with the
upper "T" denoting matrix transposition), and $g_{DSR}(E)$ is the $4\times 4$
matrix (10).

\section{DSR and Lorentz Invariance Breakdown}

Let us now discuss the link between DSR and the violation of local Lorentz
Invariance (LLI).

Theoretical speculations on the validity of LLI and SR can be traced back to
the mid of the XX century. These early works were based on the existence of
an absolute object in vacuum (like e.g. an universal length) [13].

In recent times, there has been an increasing interest in theoretical
formalisms admitting for LLI violation [14].

They can be roughly divided in two classes: unified theories and theories
with modified spacetimes. To the former one belong e.g. Grand-Unified
Theories, (Super) String/Brane theories, (Loop) Quantum Gravity, and the
so-called ``effective field theories''. The latter include \textit{e.g.}
foam-like quantum spacetimes, spacetimes endowed with a nontrivial topology
or with a discrete structure at the Planck length, $\kappa $-deformed Lie
algebra noncommutative spacetimes (for instance Doubly Special Relativity
[12]).

Although LLI breakdown has been also discussed within the framework of the
Standard Model \ [15], an extension of the Standard Model has been proposed
[16] by assuming that the breakdown of Lorentz and/or CPT invariance is due
to spontaneous symmetry breaking (namely to a non-invariance of the vacuum
under these symmetries).

More recently, it was shown by Bogoslovsky [17] that Lorentz symmetry
violation might occur without violation of relativistic symmetry represented
in such a case by the 3-parameter group of the so-called generalized Lorentz
boosts. This group serves as a \ noncompact homogeneous subgroup of the
8-parameter isometry group of the flat Finslerian spacetime with partially
broken 3D isotropy. Such event space, generalizing the Minkowski space,
coincides with that firstly introduced by the same author [8]. From this
work it follows, among the others, that the physical carrier of the
space-time anisotropy is the anisotropic fermion-antifermion condensate,
arising from spontaneous breaking of initial gauge symmetry (for instance,
in the Standard Model). Later on the results obtained in this work were
mostly reproduced with the help of the techniques of continuous deformations
of the Lie algebras and nonlinear realizations [18]\footnote{%
However, in [18] a different notation was used in comparison with [17]. In
particular, the parameter that characterizes the space anisotropy magnitude
was designated by ``$b$'' instead of~``$r$''.}. Such a formalism was called
in [18] ``General Very Special Relativity'' [19], while the 8-parameter
group of Finslerian isometries was called DISIM$_{b}$(2), i.e. Deformed
Inhomogeneous SIMilitude group, that includes the 2-parameter Abelian
homogeneous noncompact subgroup. The DISIM$_{b}$(2) invariant nonlinear
Dirac equation was proposed in ref.[20].

Coming again to DSR, let us remark the mathematically self-evident, but
physically basic, point that the generalized metric (10) (and the
corresponding interval) is clearly \emph{not preserved} \emph{by the usual
Lorentz transformations}. If $\Lambda _{SR}$ is the $4\times 4$ matrix
representing a standard Lorentz transformation, this amounts to say that the
similarity transformation generated by $\Lambda _{SR}$ does not preserve the
deformed metric tensor $g_{DSR}$:
\begin{equation}
\left( \Lambda _{SR}\right) ^{T}g_{DSR}\Lambda _{SR}\neq g_{DSR}.
\end{equation}%
This is by no means an unexpected result, at the light of the axiomatic
formulation of Special Relativity (see Sect.2). However, as a consequence,
the deformed metric structure of $\widetilde{M}$ \emph{violates} the
standard Lorentz invariance, characteristic of the usual Minkowski
space-time $M$. In this sense, therefore, we can state that DSR is strictly
related to (and able to describe) the possible breakdown of Lorentz
invariance, since the deformed metrics are no longer kept invariant by the
standard Lorentz transformations.

However, it possible to construct \emph{deformed Lorentz transformations}, (%
\textit{i.e.} isometries of $\widetilde{M}$) which \emph{do preserve} the
generalized metric and interval \linebreak (10,11)\footnote{%
For their explicit form we refer the reader to refs.[6].}. Therefore,
Lorentz invariance, broken by the energy-dependent deformation of the space
time \emph{in its usual sense}, namely as a special-relativistic symmetry
property of the interactions and/or the physical systems, \emph{is recovered}%
, in the framework of DSR, in a generalized, wider meaning. We shall name
\emph{deformed Lorentz invariance (DLI)} this extended LI.

The mathematical formulation of DLI is provided by the following equation
\begin{equation}
\Lambda _{DSR,int.}^{T}(E)g_{DSR,int.}(E)\Lambda
_{DSR,int.}(E)=g_{DSR,int.}(E).
\end{equation}%
(where we emphasized the dependence of the deformed Lorentz transformations
on the interaction considered). It can be read as follows:\smallskip

\emph{- For every physical interaction,which affects the space-time geometry
by deforming it in a way described by the metric tensor }$g_{DSR.int.}$\emph{%
, it is always possible to find deformed Lorentz transformations }$\Lambda
_{DSR,int.}$ \emph{preserving the deformed geometrical structure of
space-time for the interaction considered, }namely (from a mathematical
point of view) generating similarity transformations which leave the
deformed metric tensor invariant.

Then, we can state that DSR not only permits to deal with LI breakdown on a
physical basis, but allows one to recover Lorentz invariance as an extended,
higher symmetry of physics, valid for systems and/or interactions violating
LI according to the usual Special Relativity, in the usual Minkowski
space-time.

In conclusion, we want to stress that the DSR formalism has a number of
possible implications and developments, both from a theoretical and an
experimental view. These are, among the others: the possibility of getting
phenomenological metrics (derived from experimental data) for the four
fundamental interactions, which evidence departures from Minkowski metric in
suitable energy ranges; the extension of the formalism to a five-dimensional
scheme of the Kaluza-Klein type; the prediction of new physical effects some
of which experimentally verified [21], like the existence of piezonuclear
reactions, namely nuclear reactions triggered by pressure in liquids and
solids in non-Minkowskian conditions. We refer the reader to ref.[6] for
further details.

\label{lastpage-01}


\begin{thebibliography}{99}
\bibitem{1} B. Finzi: ``Relativit\`{a} Generale e Teorie Unitarie'', in
\textit{Cinquant'anni di Relativit\`{a}} (in Italian), ed. M. Pantaleo
(Giunti, Firenze, Italy, 1955), pg. 194.

\bibitem{2} E. Recami and R. Mignani: \textit{Riv. Nuovo Cimento} \textbf{4}%
, n.2 (1974), and references therein.

\bibitem{3} R. Penrose: \textit{The Emperor's New Mind} (Oxford University
Press, 1989).

\bibitem{4} J. Ehlers, F.A.E. Pirani and A. Schild, in \textit{Papers in
Honour of J. L. Synge}, edited by L. O'Raifeartaigh (Clarendon Press, Oxford
1972).

\bibitem{5} J.S. Anandan: in \textit{Potentiality, Entanglement and
Passion-at-a-distance - Quantum Mechanical Studies for Abner Shimony}, vol.
2, edited by R. S. Cohen, M. Horne and J. Stachel (Kluwer, Dordrecht,
Holland 1997), p. 31.

\bibitem{6} F. Cardone and R. Mignani: \textit{Energy and Geometry - An
Introduction to Deformed Special Relativity} (World Scientific Series in
Contemporary Chemical Physics, vol. 22) (World Scientific, Singapore, 2004);
\textit{Deformed Spacetime - Geometrizing Interactions in Four and Five \
Dimensions }(Springer-Verlag, New York, 2007); and references therein.

\bibitem{7} For a review of Finsler geometry, see \textit{e.g.} Z. Shen:
\textit{Lectures on Finsler Geometry }{(World Scientific, Singapore, 2001).}

\bibitem{8} G.Yu. Bogoslovsky: \textit{Nuovo Cimento B }\textbf{40}, 99
(1977). For a review, see G.Yu. Bogoslovsky: \textit{Fortsch. Phys}. \textbf{%
42}, 2 (1994).

\bibitem{9} The Lie-isotopic generalization of Special Relativity was
introduced by R.M. Santilli in \textit{Lett. Nuovo Cim. }\textbf{37}, 565
(1983). For a review of Lie-isotopic theories, see e.g. R.M. Santilli:
\textit{Found.Phys.} \textbf{27}, 625 (1997), and refs. therein..

\bibitem{10} H. Nielsen and I. Picek: \textit{Nucl.Phys.} \textbf{B211}, 269
(1983).

\bibitem{11} See \textit{e.g.} A. U. Klimyk and K. Schmudgen: \textit{%
Quantum Groups and Their Representations} (Texts and Monographs in Physics)
(Springer-Verlag, New York, 1997).

\bibitem{12} See \textit{e.g. }G. Amelino-Camelia: \textit{\ Int. J. Mod.
Phys.} \textbf{D11}, 1643 (2002), and refs. therein.

\bibitem{13} For a review of these early attempts, see \textit{e.g.} F.
Cardone and R. Mignani: \textit{Found. Phys. }\textbf{29}, 1735 (1999).

\bibitem{14} For a review, see D. Mattingly: ``Modern Tests of Lorentz
Invariance'', \textit{Living Rev. Relativity}, 8, 5 (2005) [Online Article]:
cited [7 Sept. 2005], http://www.livingreviews.org/lrr-2005-5.

\bibitem{15} S. Coleman and S.L. Glashow: \textit{Phys. Rev. D} \textbf{59},
116008 (1999); R. Jackiw: ``Chern-Simons violation of Lorentz and PCT
symmetries in electrodynamics'' (hep-ph/9811322),1998, and references
therein.

\bibitem{16} See \textit{e.g.} \textit{CPT and Lorentz Symmetry I, II, III},
V.A. Kostelecky ed. (World Scientific, Singapore, 1999, 2002 ,2004), and
refs. therein.

\bibitem{17} G.Yu.Bogoslovsky: \textit{Phys. Lett.A}, \textbf{350}, 5
(2006); \textit{SIGMA}, \textbf{1, }017\textbf{\ }(2005).

\bibitem{18} G.Gibbons, J.Gomis and C.Pope: \textit{Phys. Rev.D} \textbf{76,
}081701(R) \textbf{\ }(2007).

\bibitem{19} In this connection, see also A.G.Cohen and S.L.Glashow: \textit{%
Phys. Rev. Lett. }\textbf{97}, 021601 (2006).

\bibitem{20} G.Yu.Bogoslovsky and H.F.Goenner: \textit{Phys. Lett.A},
\textbf{323, }40 (2004).

\bibitem{21} U. Bartocci, F. Cardone and R. Mignani: \textit{Found. Phys.
Lett.} \textbf{14}, 51 (2001); F. Cardone, R. Mignani, W. Perconti and R.
Scrimaglio: \textit{Phys. Lett. A} \textbf{326}, 1 (2004); F. Cardone, R.
Mignani and A. Petrucci: \textit{Phys. Lett. A} \textbf{373}, 1956 (2009);
F. Cardone, G. Cherubini, R. Mignani, W. Perconti, A. Petrucci, F. Rosetto
and G. Spera: arxiv.org/abs/0710.5115 (Annales Fond. de Broglie, in press);
F. Cardone, G. Cherubini and A. Petrucci: \textit{Phys. Lett. A} \textbf{373}%
, 862 (2009); F. Cardone, A. Carpinteri and G. Lacidogna: \textit{Phys.
Lett. A} \textbf{373}, 4158 (2009); F. Cardone, R. Mignani and W. Perconti:
\textit{New Advances Phys.} \textbf{3}, 7 (2009).
\end{thebibliography}
\end{document}